# AGENTS AND OWL-S BASED SEMANTIC WEB SERVICE DISCOVERY WITH USER PREFERENCE SUPPORT

Rohallah Benaboud[1], Ramdane Maamri[2] and Zaidi Sahnoun[3]

[1]Department of Computer Science and Mathematics, University of Oum El Bouaghi, Algeria

r_benaboud@yahoo.fr

[2,3] LIRE Laboratory, University of Constantine 2, Algeria

[2]rmaamri@yahoo.fr, [3]sahnounz@yahoo.fr

## ABSTRACT

*Service-oriented computing (SOC) is an interdisciplinary paradigm that revolutionizes the very fabric of distributed software development applications that adopt service-oriented architectures (SOA) can evolve during their lifespan and adapt to changing or unpredictable environments more easily. SOA is built around the concept of Web Services. Although the Web services constitute a revolution in Word Wide Web, they are always regarded as non-autonomous entities and can be exploited only after their discovery. With the help of software agents, Web services are becoming more efficient and more dynamic.*

*The topic of this paper is the development of an agent based approach for Web services discovery and selection in witch, OWL-S is used to describe Web services, QoS and service customer request. We develop an efficient semantic service matching which takes into account concepts properties to match concepts in Web service and service customer request descriptions. Our approach is based on an architecture composed of four layers: Web service and Request description layer, Functional match layer, QoS computing layer and Reputation computing layer.*

## 1. INTRODUCTION

Service Oriented Architecture (SOA) is an approach to build distributed systems that deliver application functionality as services which are language and platform independent [1]. A Web service is a technology that realizes the SOA [2]. At present, many corporations and organizations have implemented their core application through buying the Web Services on Internet. For example, salesforce.com provides ERP service for users [3].

Web service has been defined as modular, self-describing, loosely-coupled application that can be published, located and invoked through the Internet. The wide spreading of Web Services is due to its simplicity and the data interoperability provided by its components namely XML (eXtended Markup Language), SOAP (Simple Object Access Protocol), UDDI (Universal Description, Discovery and Integration) and WSDL (Web Service Description Language). XML is used to describe the data format of the Web service in a structured way; SOAP is used to transfer the data; UDDI is used for discovery of services; WSDL is used for describing the services.

Web services are useless if they cannot be discovered. So, discovery is the most important task in the Web service model [4]. Web service discovery is a process of discovering service that most suitable to service consumer request according to consumer requirement. There are two challenges facing the practicality of Web services discovery [5]: (a) efficient location of the Web service registries that contain the requested Web services and (b) efficient retrieval of the best requested services from these registries.

Traditional UDDI supports only keyword-based discovering mechanisms. Keywords are insufficient in expressing semantic concepts and semantically different concepts could possess identical representation, which will further lead to low precision [6]. Therefore, several approaches have been proposed to add semantics to Web Services descriptions to facilitate discovery and selection of relevant Web services (e.g. DAML-S [7], WSDL-S [8], WSML [9], OWL-S [10]). These so-called Semantic Web services can better capture and disambiguate the service functionality, allowing a logic-based matchmaking to infer relationships between requested and provided service parameters [11]. Ontologies are a key technology to achieve Semantic Web [12]. They aim to provide a machine-processable semantics of information sources that can be communicated between different applications. They also play a very important role in the development of intelligent systems.

If the discovery engine returned multiple candidate Web services provide the same functionality, then Quality of Service (QoS) is becoming an important criterion for selection of the best available Web service [1]. As users are not clearly aware of the functional and non-functional QoS information of the existing web services, it is usually quite confusing for them to make a decision in choosing an appropriate service matching their requirements. So it is very important to extend the existing model to support QoS of web service and establish a web service evaluation approach to solve the problem. Meanwhile user preference is also important in web service selection in that a web service which has the best evaluation value is not necessarily always matching user individual requirements [13]. So when we design the web service discovery approach, we must take user preference into account simultaneously as a key factor.

Our work aims to provide a more "consumer-centric" approach simplifying service discovery using semantics while satisfying QoS requirements. To incorporate service QoS information in service discovery, the major problem is the specification and storage of the QoS information [14]. In this paper, we propose an ontology-based OWL-S extension to adding QoS to Web service descriptions.

Consumers want to select the service providers which honestly offer the service with advertised QoS. Making use of reputation, consumers can find trusted Web Services. So, the service quality's reputation is vital important to select the genuine service required by the consumers. After using a web service, the service consumer has to rate the various QoS attributes of this Web service. Each service consumer has its own satisfaction criteria to judge a Web service as good or not. For example, a service consumer may give a low rating to a Web service that has execution price more than 0.5 EUR/Sec. If the execution price is not significant for a second service consumer, then the first service consumer's low rating will not be significant either. Hence, without knowing the intendment of the service consumer, it is almost impossible to make sense of a given rating.

In this paper we present a useful approach for Web services discovery based OWL-S, software agents and domain ontologies. This approach is based on an architecture composed of four layers: Web service and Request description layer, Functional match layer, QoS computing layer and Reputation computing layer. Each layer uses the result of the previous layer and aims to decrease the number of candidate web services. Web service discovery process became more efficient and more dynamic by exploiting the parallelism and the distribution given by agent technology.

The rest of this paper is organized as follows: Section 2 surveys related research. Section 3 presents our approach for Web service discovery. In Section 4, we present how to implement the approach using software agents and section 5 offers concluding remarks and future directions.

## 2. RELATED WORKS

The researches in Web services discovery have been necessary since the number of available services on internet has increased and the customer gets tired to find desired service. Many researches have investigated the discovery of semantic web services, QoS-aware discovery or the use of agents for semantic web services discovery. In this section, we present and analyse the related work in Web services discovery in order to comprehend the benefits that may be obtained and to put our contributions in context.

### Semantic Web Services Discovery

Most current approaches for web service discovery initiative to add semantics to support Web service discovery. Especially, semantic web service discovery aims to discover the best matched web service, it mostly depends on the measurement of the similarity degrees between service request and service advertisement. The work presented in [15] proposes a DAML-S matchmaking algorithm which is used to match a requested service with a set of advertised ones. This matching algorithm compares the input and output concepts of user request to the service description in registry and defines four levels of matching: Exact, Plug in, Subsumes, Fail. In our work, we don't use only the subsumption relationships between concepts to calculate their similarity but we also take into account common properties between them.

In [12], authors propose a web service discovery method based on the domain ontologies. The proposed method calculates semantic similarity based on ontology distance between concepts in both service request and service advertisement. The web service discovery algorithm in this paper only calculates the semantic similarity of the functional description of the web service, and jumps over the semantic similarity of the description of the input/output parameters.

### QoS-aware discovery and reputation systems

Since there are many functionally similar Web Services available in the Web, it is an absolute requirement to distinguish them using a set of non-functional criteria such as Quality of Service (QoS).

According to [16], the Web service discovery is usually driven only by functional requirements, which can't guarantee the real-time validity of the web services selection. So he proposed a QoS based Web Services Selection Model (WSSM-Q) to provide QoS support for service publishing and selection. In the model, the QoS of web services is managed, including defining the QoS model, collecting the QoS information, computing and maintaining the QoS data. Upon this QoS management, the web services that match the requirements of consumers are ranked for selection according to the overall QoS utility. The author in [17] proposes a QoS-based model for web service discovery by extending the UDDI's data structure types in order to enhance UDDI model with QoS information. However service discovery and selection are still done by human consumer. This approach is impractical with a huge number of Web services available for selection. The work described in [18] refers to the need for an extensible QoS model that contains domain-specific QoS criteria. It sustains that QoS must be represented to consumers according to consumer preferences and consumers should express accurately their preferences with this QoS model without resorting to complex coding of consumer profiles. It also suggests that QoS computation must be fair and open for providers and requesters. Then it proposes an extensible QoS model.

Some existing researches are variants of reputation systems based on the feedback obtained from service consumers. In [19], authors had proposed "RATEWeb", a framework for establishing trust in service-oriented environments. A cooperative model is used in which Web services share their experiences of the service providers with their peers through feedback ratings. The characteristic of this framework was the credibility of the consumers of evaluating services has been taken into account. If the rater tried to provide a fake rating, then its credibility would be decreased and the rating of this user would become less important in the reputation of the web service. The work presented in [14] proposes a model of reputation-enhanced Web services discovery with QoS. The proposed model contains an extended UDDI to support the QoS information which is expressed in XML style format and stored using tModels in a UDDI registry. It contains also a reputation manager which ranks services that meet a customer's functionality and QoS requirements using the service reputation scores based on user feedback regarding their performance. In this work, the satisfaction criteria of the rater is unknown since the service consumer gives one rating score for all QoS of the invoked service.

**Agents and Web Services Discovery**

The use of software agents in Web Service discovery has been the subject of many researches. Authors in [20] proposed a multi-agent approach to achieve service discovery and selection according to the consumers' preferences. They describe a system in which agents interact and share information, in essence creating an ecosystem of collaborative service providers and consumers. In [21], authors propose a multi-agent approach for a distributed information retrieval task. In this work, each agent has a view of its environment called agent view. The agent-view structure of an agent contains information about the language models of documents owned by each agent. An agent-view reorganization algorithm is run to dynamically reorganize the underlying agent-view topology. The proposed protocol does not use ontologies during information retrieval.

## 3. A QoS AND CONSUMER PREFERENCE BASED APPROACH FOR WEB SERVICE DISCOVERY

Our proposed approach finds similar services to the consumer request based on functional and QoS similarity, and reputation computing. This approach is based on an architecture composed of four layers (Figure 1). Each layer uses the result of the previous layer (just below) and aims to decrease the number of candidate web services.

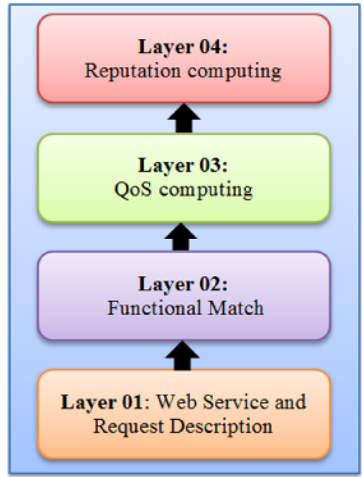

Figure 1. Approach architecture

We briefly explain what happens in each layer as follows:

– In the Web service and Request description layer, the service providers register their Web services and provide functional and non-functional information about the offered services. The service consumer inserts his request with functional properties and QoS preferences. The result of this phase is a request description file and web description file enhanced with QoS parameters. OWL-S approach is used to describe both request and web services parameters

– In the Functional match layer, functional properties of request and web service descriptions are matched based on syntactic and semantic match. The result of this phase is a set of candidate services which match the request functional parameters.

– In the QoS computing layer, we calculate the QoS score of each candidate web service and eliminate web services that does not meet the constraints or the preference of web service requester.

– In the Reputation computing layer, we calculate the reputation score of each candidate web service and then services are sorted according to overall score witch is the sum of functional similarity, QoS score and reputation score.

## 3.1. Web service and Request description layer

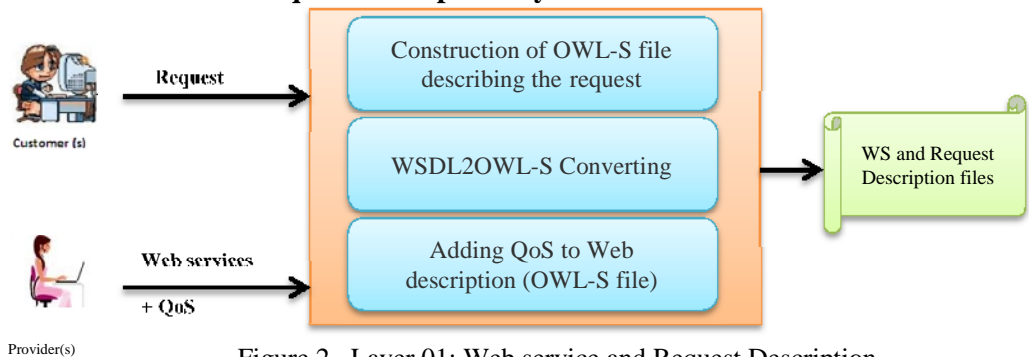

Figure 2. Layer 01: Web service and Request Description

In our work, both web service and request are described using OWL-S approach. OWL-S is an OWL-based Web service ontology to facilitate the automation of Web service tasks, including automated Web service discovery, execution, composition and interoperation through their semantic description [10]. The OWL-S ontology has three main parts:

- − The service profile that describes what the service does.
- − The process model that specifies how the service works.
- − The service grounding that defines how the service is accessed.

If the web service provider publishes web services as Web Service Description Language (WSDL) which syntactically describes a web service, then it is converted in to OWLS description by using OWL-S plug-in available with Protégé 2000.

Typically, Web services are described using functional and non-functional properties. Functional properties of services represent the description of the service functionalities. In our work, functional properties contain Service Name, Textual description, a set of Inputs and a set of Outputs. Non-functional properties represent QoS like response time, reliability, availability, security and execution cost and so on. We use OWL-S service profile as a model for semantic matchmaking of service descriptions. However, OWL-S mainly focuses on describing functional aspects of a Web service. Based on works presented in [22] [23], we propose a simple ontology-based OWL-S extension to adding non-functional description, referring to as QoS, to Web service description. The new service profile model is depicted in Figure. 3. In OWL-S service profile we can use a set of ServiceParameter which has name (serviceParameterName) and a value (sParameter). For the connection of OWL-S and QoS ontology, the QoSProperty is a subclass of OWL-S ServiceParameter. And QoSParameterName and qosParameter are subproperties of OWL-S ServiceParmaerterName and sParameter property.

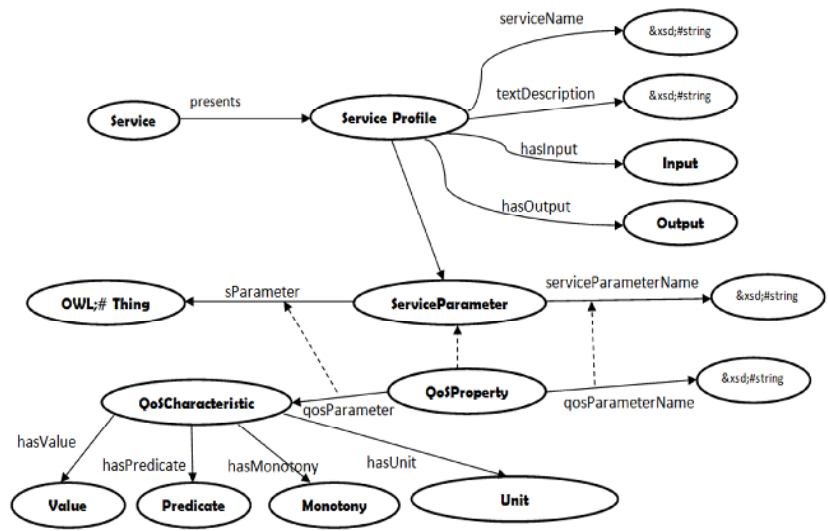

Figure 3. OWL-S extension to support QoS

We give an example of a specific QoS profile as follows:

```
< qosProf:QoSPrpoperty rdf:ID=" QoSParamIndividual">
< qosProf:qosParameterName rdf:datatype= "&xsd string;   ResponseTime"/>
<qosProf:QoSCharacteristic rdf:ID=" QoSCharctiristicIndividual">
< qosProf:hasValue rdf:datatype="&xsd integer">600 </hasValue>
< qosProf:hasPredicate rdf:resource="&qos;less"/>
< qosProf:hasUnit rdf:resource="&qos;millisecond"/>
< qosProf:hasMonotony rdf:resource="&qos;decrease"/>
</ qosProf: QoSCharacteristic >
</qosProf: QoSPrpoperty>
```

A request signifies a service demand. Request description includes functional and non-functional requirements. The former describes the functional characteristic of the service demand, such as name, textual description, inputs and outputs. The latter mainly focuses on the customer's preferences, namely quality of service (QoS).

Service consumers have different preferences. For example, a service consumer may want a Web service with lower cost while for another response time could be his most important parameter. For this raison, we propose that service consumer doesn't have to give the value of each desired QoS attribute but he should get the means to specify that a QoS attribute is more important than another one. Indeed, he gives a weight for each QoS attribute. Weights range from 0 to 5 where higher weights represent greater importance. Figure 4 shows an example of a user request interface with weights of some QoS attributes.

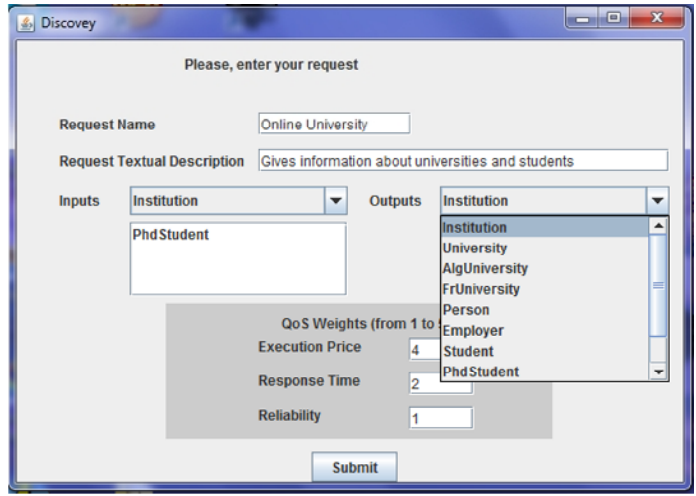

Figure 4.  Consumer Request Interface

## 3.2. Functional match Layer

In this layer, we take functional properties of request and web service then we calculate the similarity between them. Name and textual description of request and services are matched using syntactic similarity function whereas inputs and outputs are matched based on semantic similarity function.

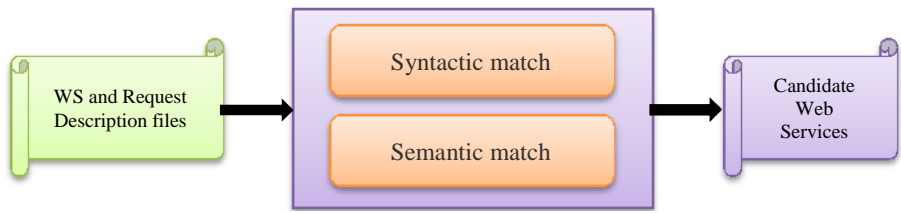

Figure 5. Layer 02 : Functional match

### 3.2.1. Syntactic Similarity

In our work, we use the q-grams methodology to evaluate syntactic similarity. Q-grams methodology [24] takes as input two strings and computes the number of common q-grams between them. Q-grams are sequences of *q* characters. These subsets are called grams and the quantity of characters in each gram is defined by *q*. After some experiments we chose to use a 3-gram comparison.

To calculate the syntactic similarity, we use the algorithm 1, where the function Subset(String) takes as input a string and devises it into sequences of three characters. The function card(L) returns the number of elements in the set L. In line 4, we normalize the SyntacticSim value to have a value in the range of 0 to 1.

| **Algorithm 1** SyntacticSim(String1, String2) |
| --- |
| L1, L2, L3: sets of string; |
| **Begin** |
| 1: L1 = Subset(String1); |
| 2: L2 = Subset(String2); |
| 3: L3 = L1 ∩L2; |
| 4: SyntacticSim = $\frac{2 \cdot card(L3)}{card(L1) + card(L2)}$ ; |
| 5: **return** SyntacticSim; |
| **End** |

Example: Table 1 shows how to calculate the SyntacticSim between "FindAlgUniversity" and "FindAlgerianUniversity".

Table 1. Example of SyntacticSim

| | FindAlgUniversity | FindAlgerianUniversity |
| --- | --- | --- |
| **3-grams** | fin, ind, nda, dal, alg, lgu, gun, uni, niv, ive, ver, ers, rsi, sit, ity | fin, ind, nda, dal, alg, lge, ger, eri, ria, ian, anu, nun, uni, niv, ive, ver, ers, rsi, sit, ity |
| **SyntacticSim** | $\frac{2 \cdot 13}{15+20}$ = 0,743. | |

## 3.2.2. Semantic Similarity

We consider that all inputs and outputs refer to concepts of domain ontology. In fact, matching inputs (outputs) of the request and the service is nothing other than the matching of concepts associated to inputs (outputs). To calculate the similarity of two concepts X and Y, we distinct many scenarios:

- The concepts are same or they declared as equivalent classes (X=Y).
- The concept X is subclass of the concept Y directly or indirectly (X<Y).
- The concept Y is subclass of the concept X directly or indirectly (Y<X).
- X does not have a parent/child relationship with Y, but both concepts have a parent concept Z in common (X<>Y).
- Otherwise (X≠Y)

As it is shown in [25], an output in the request must not be consider as similar to a more generic output in the advertised service, while a request input could be consider as similar to a more generic advertised input. We think also that an input in the advertised service must not be consider as similar to a more generic input in the request, while an output in the advertised service could be consider as similar to a more generic output in the request.

The function nbprop(X) denotes the number of properties of the concept X. The intersection X∩Y denotes the intersection of the set of properties in X and the set of properties in Y.

ConceptSim(X, Y) matches the concept X (Y ∈ request inputs or outputs) against the concept Y(Y∈ service inputs or outputs) and returns the conceptual similarity of the two concepts. All inputs and outputs refer to concepts of domain ontology, an example portion of which is shown in figure 6 in a logic description notions. we give in Table 3 the exact definition of the function ConceptSim(Y, X).

Table 3. ConceptSim Definition

| | Concepts are Inputs or Outputs in Req and WS | Relationship between Concepts in Domain Ontology | ConceptSim(X, Y)= |
|---|---|---|---|
| **Case 01** | X, Y are Inputs or Outputs | X=Y | 1 |
| **Case 02** | X, Y are Inputs | X<Y | 1 |
| **Case 03** | X, Y are Inputs | Y<X | $\dfrac{\text{nbprop}(X)}{\text{nbprop}(Y)}$ |
| **Case 04** | X, Y are Outputs | X<Y | $\dfrac{\text{nbprop}(Y)}{\text{nbprop}(X)}$ |
| **Case 05** | X, Y are Outputs | Y<X | 1 |
| **Case 06** | X, Y are Inputs or Outputs | X<>Y | $\dfrac{\text{nbprop}(X \cap Y)}{\text{nbprop}(X \cup Y)}$ |
| **Case 07** | X, Y are Inputs or Outputs | X≠Y | 0 |

University ≡ Institution ∩(∀hasID.UniversityID) ∩ (=1hasID) ∩(∀hasName.Name)
    ∩(=1hasName) ∩ (∀hasPostcode.Postcode) ∩ (=1hasPostcode) ∩
    (∀hasCourse.Course) ∩ (=1hasCourse)

AlgUniversity ≡ University ∩(∀hasPostcode.AlgPostcode) ∩ (=1hasPostcode)

Person ≡ (∀hasAdress.Adress) ∩ ( 1hasAdress) ∩ (∀hasFirstName.Name) ∩
    (=1hasFirstName) ∩(∀hasLastName.Name) ∩ (=1hasLastName)

Employer ≡ Person ∩ (∀hasEmployerID. EmployerID) ∩ (=1hasEmployerID)

Student ≡ Person ∩ (∀hasStudentID. StudentID) ∩ (=1hasStudentID)

PhdStudent ≡ Student ∩ (∀hasThesisID. ThesisID) ∩ (=1hasThesisID)

GeographicArea ≡ (∀hasGoName.Name) ∩ (=1 hasGoName) ∩
    (∀hasCountryName.Name) ∩(=1 hasCountryName)

Location ≡ GeographicArea ∩ (∀hasAltitude. Altitude) ∩ (=1 hasAltitude) ∩
    (∀hasLatitude. Latitude) ∩ (=1 hasLatitude) ∩ (∀hasLongitude. Longitude)
    ∩ (=1 hasLongitude)

Figure 6. Part of sample AL Ontology

Example: For illustration, let us take the tow requests and two web services described in Table 4. All inputs and outputs refer to concepts of domain ontology shown in Figure. 6.

Table 4. Example of ConceptSim

| Request/ Web Service | Inputs | Outputs |
|---|---|---|
| **Req1** | Phstudent | Location, AlgUniversity |
| **WSer1** | Person | Location, University |
| **Req2** | GeographicArea, Person | University |
| **WSer2** | Location, Phstudent | AlgUniversity |

The different cases shown in Table 3 can be illustrated as follows:

- Case 01: ConceptSim(University, University) = 1.
- Case 02: ConceptSim(PhdStudent, Person) = 1.
  Where PhdStudent ∈ Req1.Inputs and Person ∈ WSer1.Inputs
- Case 03: ConceptSim(AlgUniversity, University) = 0,80.
  Where AlgUniversity ∈ Req1.Outputs and University ∈ WSer1.Outputs
- Case 04: ConceptSim(University, AlgUniversity) = 1.
  Where University ∈ Req2.Outputs and AlgUniversity ∈ WSer2.Outputs
- Case 05: ConceptSim(Person, PhdStudent) = 0,60.
  Where Person ∈ Req2.Inputs and PhdStudent ∈ WSer2.Inputs
- Case 06: ConceptSim(PhdStudent, Employer) = 0,50.
- Case 07: ConceptSim(Person, University) = 0.

After describing the conceptual similarity, we give now the algorithm of inputs matching. Where R.Inputs and S.Inputs denote the set of inputs in the request R and the set of inputs in the service S respectively, Card(E) denotes the cardinality of the set E, Sort(A) allow to sort the elements of the array A in descending order. In lines 1, 2, 3 and 4, the algorithm matches each request input against all Web service inputs, and keeps the best mapping for each request input. In the number of request inputs is less than the number of service inputs (line 9) then we have a miss of information, therefore InputsSim value is decreased (line 10).

---

**Algorithm 2** InputsSim(R.Inputs, S.Inputs)

InSim: array of float;
   **Begin**
1:  **foreach** $e_1$ in R.Inputs **do**
2:     **foreach** $e_2$ in S.Inputs **do**
3:      InSim$_i$ =Max(InSim$_i$ ,ConceptSim($e_1$, $e_2$));
4:     **end for**
5:     i = i + 1;
6:  **end for**
7:  *Sort*(InSim);
8:  $m$ = Card(R.Inputs) – Card(S.Inputs);
9:  **if** m<0 **then**
10:    InputsSim = $\frac{\sum_{j=1}^{Card(R.Inputs)} InSim_i}{Card(R.Inputs)}/(|m| + 1)$
11:  **else**
12:    InputsSim = $\frac{\sum_{j=1}^{Card(S.Inputs)} InSim_i}{Card(S.Inputs)}$
13:  **end if**
14:  **return** InputsSim
**End**

---

The outputs similarity given by OutputsSim(R.Outputs, S.Outputs) function is also calculated in the same way as inputs similarity. But when the number of service outputs is less than the number of request outputs, the value of OutputsSim is decreased. Therefore we inverse line 10 with 12 and perform changes in variable names in the algorithm 2.

Example: let us calculate the Inputs and Outputs similarity between req1 and WSer1 shown in previous example.

InputsSim = ConceptSim(PhdStudent, Person) = 1.
OutputsSim= $\frac{ConceptSim(Location,Location) + ConceptSim(AlgUniversity,University)}{2} = \frac{1+0,8}{2} = 0,9$.
Based on InputsSim and OutputsSim, we can calculate the semantic similarity between a request R and a web service S as bellow:

$$\text{SemanticSim(R, S)} = \frac{\text{InputsSim(R.Inputs, S.Inputs)} + \text{OutputsSim(R.Outputs, S.Outputs)}}{2};$$

Finally, functional similarity can be calculated using algorithm 3. Where weights w1 and w2 are real values between 0 and 1 and must sum to 1; they indicate the degree of confidence that the service consumer has in the syntactic similarity and semantic similarity.

---

**Algorithm 3** FunctionalSim(R, S)

---

NSim, TDSim, NTDSim, ISim, OSim, IOSim: float;
  **Begin**
1:  NSim = SyntacticSim(R.name, S.name);
2:  TDSim =
    SyntacticSim(R.TextualDescription, S.TextualDescription);
3:  NTDSim = $\frac{NSim + TDSim}{2}$;
4:  ISim = InputsSim(R.Inputs, S.Inputs);
5:  OSim = OutputsSim(R.Outputs, S.Outputs);
6:  IOSim = $\frac{ISim + OSim}{2}$;
7:  FunctionalSim = (w1 * NTDSim) + (w2 * IOSim);
8:  **return** FunctionalSim;
  **End**

---

## 3.3. QoS computing Layer

In this layer, we manage QoS attributes and values. We take into account that each QoS attribute is monotonically increasing or decreasing. A monotonically increasing QoS attribute means increases in the value reflects improvements in the quality (ex. Reliability), while monotonically decreasing means decreases in the value reflects improvements in the quality (ex. Execution Price and Response Time) [14].

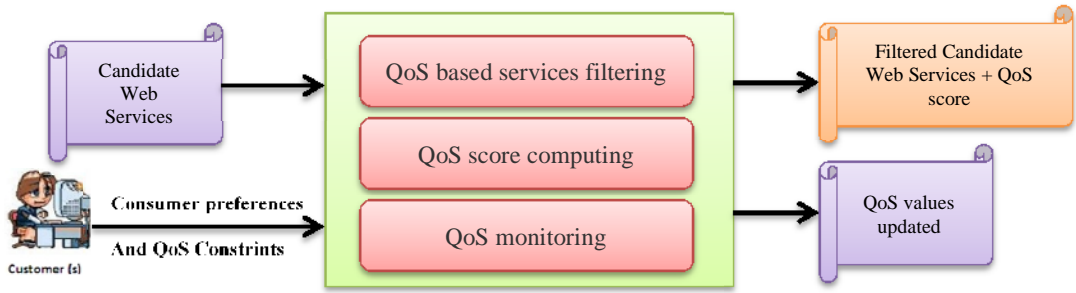

Figure 7. Layer 03 : QoS Computing

### 3.3.1. QoS based services filtering

Sometimes, the service consumer has certain minimum requirements or maximum requirements on some QoS attributes of the desired services. For example, a service consumer may want a service that offers an execution price not exceeding 100 units. So, candidate services which are over this threshold value will be eliminated. Therefore, our approach uses a QoS-based services filtering algorithm. With this algorithm we can avoid the selection of web services that does not meet the service consumer preference. Algorithm 4 takes as inputs a set of candidate services and a set of QoS based constraints and then remove unwanted services taking into account that each QoS attribute can be monotonically increasing or decreasing.

---

**Algorithm 4** QoSServicesFiltering(CandidateServices, QoSConstraints)

---

    **Begin**
1: **foreach** service S in CandidateServices **do**
2:   **if** (QoSParameter.name = QoSConstraints. name) **and**
     (QoSCharacteristic.Monotony = "increase" ) **and**
     (QoSCharacteristic.value < QoSConstraints. threshold) **then**
       Eliminate (S).
   **Endif**.
3: **if** (QoSParameter.name = QoSConstraints. name) **and**
     (QoSCharacteristic.Monotony = "decrease" ) **and**
     (QoSCharacteristic.value > QoSConstraints. threshold) **then**
       Eliminate (S).
   **Endif**.
4: **end for**
   **End**

---

### 3.3.2. QoS score computing

Each QoS value needs to be normalized to have a value in the range of 0 to 1. To normalize the QoS value, we take into account that each QoS attribute is monotonically increasing or decreasing. Monotonically increasing QoS attribute are normalized by Equations (1) and monotonically decreasing QoS attribute are normalized by Equations (2). In addition, qos.max value and qos.min value show the maximum and minimum value of the QoS attribute between all candidate services.

NoramizedValue(qos)=

$$\begin{cases} 1 - \frac{qos.\max - qos}{qos.min - qos.min} \ if(qos.\max \quad qos.\min) \\ 1 \qquad\qquad\qquad\qquad if(qos.\max = qos.\min) \end{cases} \qquad (1)$$

NoramizedValue(qos)=

$$\begin{cases} 1 - \frac{qos - qos.min}{qos.\max - qos.min} \ if(qos.\max \quad qos.\min) \\ 1 \qquad\qquad\qquad\qquad if(qos.\max = qos.\min) \end{cases} \qquad (2)$$

To calculate the overall QoS score of the service S, each normalized QoS attribute, qos is multiplied the corresponding weight, w, given by a service consumer as shown by Equation bellow:

$$QoSScore(S) = \frac{qos \quad w}{w}$$

### 3.3.3. QoS monitoring

We distinguish two types of QoS attributes: static and dynamic. Here, the static one indicates that the value of QoS attribute is known before service execution, for example, execution price. The dynamic one is for QoS attribute that is known after service execution, for example, execution time.

In the QoS monitoring step, we monitor the candidate web services to confirm accessibility, availability, and collects actual QoS values for dynamic QoS attributes.

### 3.4. Reputation computing Layer

In this layer, we collect consumer reviews on QoS of the services they had invoked. After that we calculate reputation scores based on these reviews. Finally, we store these scores in the Rating Database. To do this, we assume that all ratings are valid and objective.

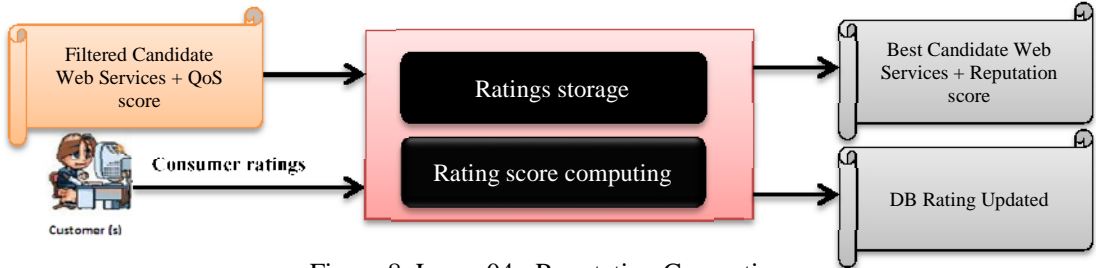

Figure 8. Layer 04 : Reputation Computing

### 3.4.1. Ratings storage

Service consumers should provide a rating score to show the level of nicety about each QoS attribute after communication. A rating score is an integer number that ranges from 0 to 5, where 5 signify most satisfaction. The consumer ratings are stored in an RDF triple store. As consumer ratings refer to a given service request, each Rating instance contains the consumer who performed the rating, the service request, the Provider who provided the service, the rated service, and finally the rating scores (one rating score per QoS). New ratings from the same consumers for the same service replace older ratings. For example, a rating from service consumer C1 (we use the IP address of the consumer) about a request R1 and a service S1 would be stored as:

```
<r:Rating>
  <r:Consumer rdf:about="#consumerC1"/>
  <r:Request rdf:about="#requestR1"/>
  <r:ProviderAgent rdf:about="#ServiceProvider"/>
  <r:Service rdf:about="#serviceS1"/>
  <r: ExecutionPriceScore rdf:datatype="&xsd;integer">2 </r: ExecutionPriceScore >
  <r: RespenseTimeScore rdf:datatype="&xsd;integer">4   </r: RespenseTimeScore >
  <r…………………………………………………………..
  </r…………………………………………………………..
</r:Rating>
```

### 3.4.2. Rating score computing

The reputation score of a service within multiple QoS attributes is computed as the weighted sum of the reputation score of each QoS attribute.

$$\mathrm{ReputationScore}(S) = \frac{\mathrm{Rate} \quad w}{w}$$

The reputation score (Rate) of a service S within a quality attribute is computed as the average of all ratings the service receives from service consumers as indicated in Equation bellow, where N is the number of ratings for the service S. Each rating score is normalized, as a monotonically increasing criterion, to have a value in the range of 0 to 1.

$$\mathrm{Rate}(S) = \mathrm{NoramizedValue}\left(\frac{S.\mathrm{getRate}()}{N}\right)$$

# 4. AGENT-BASED IMPLEMENTATION FOR THE SERVICE DISCOVERY APPROACH

In this section, we present an agent based implementation in which a team of agents, each with local information, collaborates to satisfy Web services discovery objective. Many of the complex problems are being solved by software agents because of the flexibility they offer in design and implementation. Our work combine multi-agent approach with semantic web services to enable dynamic and efficient web service discovery, thus providing consumer with relevant high level services depending on their preference.

## 4.1. General implementation architecture

The implementation has two types of agents are devised namely, Consumer Agent and Provider Agent. We use a central base of OWL Ontologies as a reference to develop the various local ontologies and semantic descriptions of the different Web services.

We use the same language of semantic descriptions (OWL-S) and a central base of ontologies in the reason to ensure the homogeneity. The main rules in our implementation, as presented in our previous work, are [26]:

– Each Provider Agent implements a number of Web services described semantically with OWL-S enhanced with QoS attributes.
– The central ontologies base will be consulted periodically by Provider Agents to develop or enrich local ontologies.
– Domains Ontologies are obtained and updated periodically using local ontologies in the reason to ensure the homogeneity.
– The register depositories contain the advertisements of the services in OWL-S format. It maintains set of files in the form of folders. For each service a separate folder is maintained that consists of OWL-S enhanced with QoS attributes.
– The central OWL ontologies base contains the different concepts used in diverse fields of proposed Web services.
– Rating Database is used as a cache memory that stores old satisfied requests to be used in future queries which allow to free consumers from time consuming human computer interactions and Web search.

Figure 9 shows the proposed implementation. In a nutshell, it works as follows:

– Initially, service provider registers the Web service with the Provider Agent and provides functional and non-functional information about the offered service.
– When a service consumer wants to insert his request, An Ontology-Guided Interface is offered by the Consumer Agent. In order to input the request, service consumer must select the desired terms they want to use in his request from the list of terms provided by the interface in a pop-up. Entries that are not in the pop-up list are not accepted by the system. This list of terms is generated by the Consumer Agent using terms in OWL domains ontologies.
– After that, Consumer Agent checks the Rating Database to find if they exist old satisfied requests similar to the new request. In this case, it suggests the corresponding Web services to the service consumer. In the other case (e.g. no old similar requests are finding) or the service consumer isn't satisfied by the suggested services, the Consumer Agent diffuses the request to all Provider Agents.

− When receiving the request from Consumer Agent, each Provider Agent matches the request to the services in the Register Depository using OWL-S description and OWL local ontologies to perform a functional and non-functional semantic match. In the end, Provider Agent returns to Consumer Agent a set of candidate services.

− When receiving all responses from Provider Agents, the Consumer Agent sorts all candidate services according to the degree of similarity, the rating score and the QoS score. Then, the Consumer Agents returns to the service consumer the services that have the highest overall score.

− After the successful invocation of the desired service, the service consumer rates each QoS attribute of the service invoked. The rating score shows the level of nicety about each QoS attribute.

− Finally, The Consumer agent subscribes the rate values in the Rating database.

− After the web service is invoked, the Provider Agent captures and updates the actual values of dynamic QoS attributes.

− If there is a change in any service description or in service QoS information, the Provider Agent informs the Consumer Agent. This latter removes changed service from the Rating database. When a service, stored in Rating database, is not used from a long period it must be removed too.

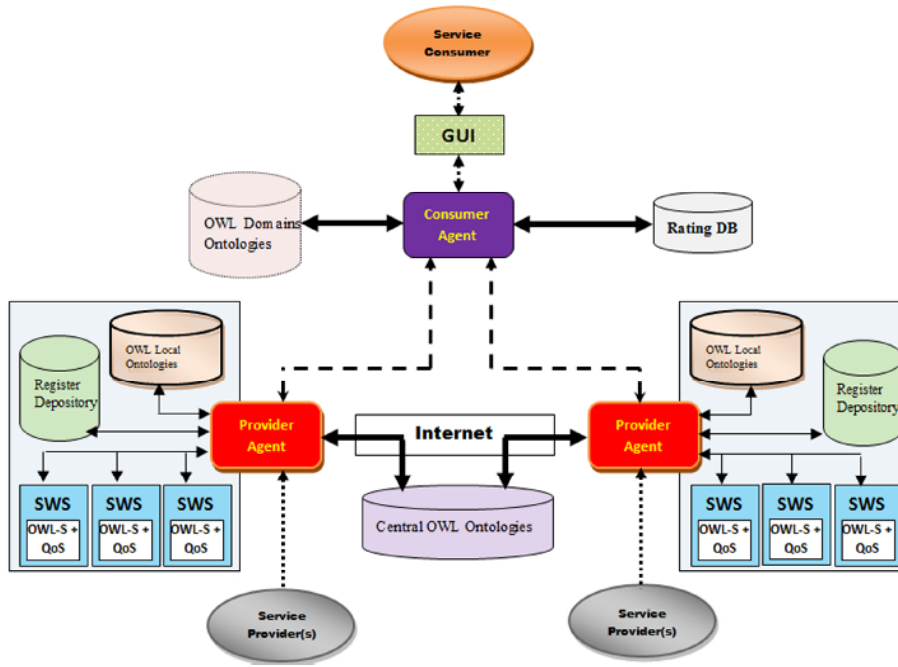

Figure 9. Agent-based implementation for WS discovery

## 4.2. Agents behavior

In the proposed implementation, each agent performs discovery when receiving a request, but there is a distinction between Consumer Agent behavior and Provider Agent behavior.

### 4.2.1. Consumer Agent behavior

When the Consumer Agent receives a request R from a service consumer, its behavior is described by algorithm 5. Where threshold is used to ignore services with a low functional similarity. CandidateServices is a local matrix where each row contains service id, functional similarity, QoS score and reputation score and overall score. Overall score is the sum of functional similarity, QoS score and reputation score. Services are sorted according to overall score.

---

**Algorithm 5: Consumer Agent Behavior 1**

---

**Begin**
**foreach** S in Rating DB **do**
   **if** FunctionalSim(R, S)   *threshold* **then**
     CandidateServices.Add(S);
   **end if**
**end for**
**if not empty** (CandidateServices) **then**
   QoSComputing(CandidateServices, QoSWeights, QoSConstraints);
   ReputationComputing(CandidateServices, QoSWeights);
   CalculateOverallScore(CandidateServices);
   Sort(CandidateServices);
   Show the result to the service consumer;
**Else**
   Diffuse the request R to all Provider Agents
**End if**
**End**

---

When Consumer Agent receives responses from all Provider Agents, it aggregates the responses as follows:

---

**Algorithm 6: Consumer Agent Behavior 2**

---

**Begin**
**foreach** received  message(MtxServices) **do**
CandidateServices.Update(MtxServices);
Sort(CandidateServices);
**End for**
**End**

---

MtxServices is a matrix where each row contains service id, functional similarity and QoS score (Initially QoS score = 0).

### 4.2.2. Provider Agent behavior

When a Provider Agent receives a request R and a threshold value from the Consumer Agent, its behavior is described as follows:

---

**Algorithm 7: Provider Agent Behavior 1**

---

**Begin**
**foreach** S in Register depository **do**
  **if** FunctionalSim(R, S)   *threshold* **then**
    CandidateServices.Add(S)
  **end if**
**end for**
**if not empty** (CandidateServices) **then**
QoSComputing(CandidateServices, QoSWeights, QoSConstraints);
CalculateOverallScore(CandidateServices);
  **End if**
Send Response(CandidateServices) to the Consumer Agents;
**End**

---

### 4.3. Testing and Programing

To program and test our approach, we have used JADE platform [27]. JADE is a very powerful middleware framework built with Java to design a MultiAgent Systems based architecture. Consumer Agent and different Provider Agents are created with JADE (Figure 10) and inherent "Agent" JADE class. Each agent has a "match module" which is realized using Jena APIs which provides plenty of methods to access ontology files. Registers Depository of our system are stored in Oracle 10g.

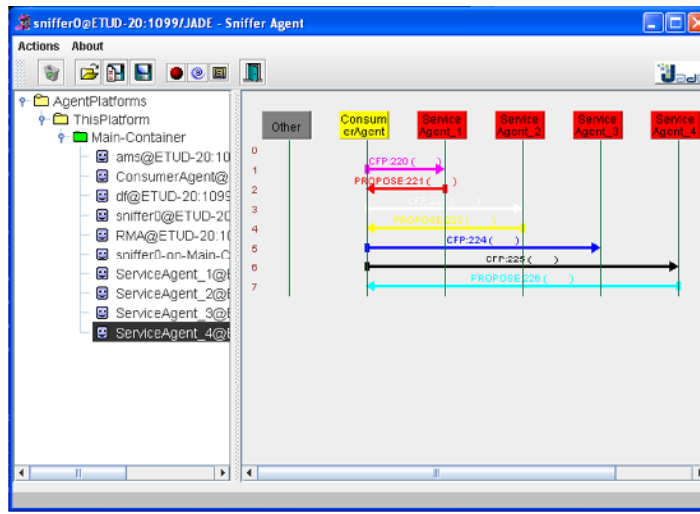

Figure 10. , JADE agents interaction

In the implementation platform, we have made some experiments to estimate the performance of our system. The Consumer Agent receives services requests with QoS weights and rating scores from a consumer simulation program. Each Provider Agent implements some Web services using OWL-S description enhanced with QoS attributes.

In our experiments, we assume that every consumer request has the same functional requirements, but different QoS weights. Inputs and outputs of request and Web services are all annotated using the ontology depicted in figure 6.

We have used four Provider Agents and each one implements five Web services. As the simulation progresses, new service ratings are generated, and the service reputation scores become different. Hence, the selected services change. After 100 runs, the experiment has illustrated that the chance of selecting a service, which better meets the QoS which a service customer was declared as important, is enhanced when previous service consumers have given a high rating score for the same QoS attribute.

## 5. CONCLUSION

This paper has introduced an approach for Web services discovery which combines multi-agent techniques with semantic web services. In the proposed approach, the service consumer has to express his preferences by specifying that a QoS attribute is more important than another one and giving the level of nicety about each QoS attribute of the service after communication. To enhance this approach, in future, we will devote into the following works:

- – Compose the functionality of several web services into one composite service to satisfy the requests when there is no single fit service.
- – Realize consumer preference and QoS-based WS composition.

**Authors**

**Rohallah Benaboud** received his BS degree in Computer Science from Mentouri University of Constantine (Algeria) in 2002, and MS degree in Computer Science from University of Oum El Bouaghi (Algeria) in 2005. He is currently a PhD student in LIRE Laboratory (Constantine 2, Algeria). He is working as Assistant professor in Department of Mathematics and Computer Science in University of Oum El Bouaghi (Algeria).

**Ramdane Maamri** is an associate Professor in TLSI Department of the University of Constantine 2, Algeria. He heads the research group 'Software Engineering and Artificial Intelligence" in LIRE Laboratory. His research interests include multi-agents systems, web applications and software engineering.